\documentclass[prb,
twocolumn,
superscriptaddress,showpacs,amsmath,amssymb]{revtex4}
\usepackage{amsfonts}
\usepackage{bm}
\usepackage{verbatim}
\usepackage{color}
\usepackage{graphicx}
\usepackage[applemac]{inputenc}

\newcommand{\tr}{\textrm{tr}}

\newcommand{\im}{i}

\newcommand{\doo}{\partial}

\begin{document}

\title{Elasticity tetrads, mixed axial-gravitational anomalies, and 3+1d quantum Hall effect}

\author{J.~Nissinen}
\affiliation{Low Temperature Laboratory, Aalto University,  P.O. Box 15100, FI-00076 Aalto, Finland}
\email{jaakko.nissinen@aalto.fi}

\author{G.E.~Volovik}
\affiliation{Low Temperature Laboratory, Aalto University,  P.O. Box 15100, FI-00076 Aalto, Finland}
\affiliation{Landau Institute for Theoretical Physics, acad. Semyonov av., 1a, 142432,
Chernogolovka, Russia}

\date{\today}

\begin{abstract}
For two-dimensional topological insulators, the integer and intrinsic (without external magnetic field) quantum Hall effect is described by the gauge anomalous (2+1)-dimensional [2+1d] Chern-Simons (CS) response for the background gauge potential of the electromagnetic U(1) field. The Hall conductance is given by the quantized prefactor of the CS term, which is a momentum-space topological invariant. Here, we show that three-dimensional crystalline topological insulators with no other symmetries are described by a topological (3+1)-dimensional [3+1d] mixed CS term. In addition to the electromagnetic U(1) gauge field, this term contains elasticity tetrad fields $E^{\ a}_{\mu}({\bf r},t) = \doo_{\mu}X^a(\mathbf{r},t)$ which are gradients of crystalline U(1) phase fields $X^a(\mathbf{r},t)$ and describe the deformations of the crystal. For a crystal in three spatial dimensions $a=1,2,3$ and the mixed axial-gravitational response contains three parameters protected by crystalline symmetries: the weak momentum-space topological invariants. The response of the Hall conductance to the deformations of the crystal is quantized in terms of these invariants. In the presence of dislocations, the anomalous 3+1d CS term describes the Callan-Harvey anomaly inflow mechanism. The response can be extended to all odd spatial dimensions. The elasticity tetrads, being the gradients of the lattice U(1) fields, have canonical dimension of inverse length. Similarly, if such tetrad fields enter general relativity, the metric becomes dimensionful, but the physical parameters, such as Newton's constant, the cosmological constant, and masses of particles, become dimensionless.
\end{abstract}
\pacs{
}

\maketitle

\section{Introduction}

The effective Chern-Simons (CS) description of the integer (and fractional \cite{ZhangHanssonKivelson89, Read89}) quantum Hall effect (IQHE) and the ensuing topological quantization of Hall conductivity has been originally considered in even spatial dimensions \cite{Thouless1982}. Similarly, in $2+1$-dimensional topological insulators \cite{Haldane1988}, in the generalized QED$_3$ \cite{Ishikawa1986,Matsuyama1987,KaplanEtAl93} or in thin films of topological superfluids and superconductors \cite{Volovik1988, Volovik92},
the intrinsic or anomalous quantum Hall effect (AQHE) in the absence of magnetic flux is also described by a CS term. For both IQHE and AQHE, the prefactor of this term is expressed in terms of momentum-space topological invariants -- the Chern number(s). The same mechanism works for gapped systems in all even space dimensions \cite{ZhangHu2001,Volovik2003, QiHughesZhang08}.

Here we show that the IQHE/AQHE in $3+1$-dimensional crystalline quantum Hall systems or topological insulators \cite{Halperin1987} is also described by CS term with mixed field content. Namely, the 3+1d CS term features elasticity tetrads
\cite{DzyalVol1980,VolovikDotsenko1979, AndreevKagan84, NissinenVolovik2018b}, which describe the geometry of elasticity theory, including crystals with dislocation defects. The density of dislocations corresponds to spatial torsion of the geometry, more familiar in gravitational theories, see e.g. Ref. \onlinecite{Kleinert}. In contrast to the gravitational tetrads, the elasticity tetrads have canonical dimensions of  inverse length. As a result the CS term is dimensionless (in units $\hbar =1$), and as in the case  of even space dimensions, the prefactor is given by integer momentum-space topological invariants.  The CS term leads to the analog of mixed (axial and gravitational/elastic) anomaly inflow in 3+1d, see e.g. Refs. \onlinecite{EguchiFreund76, CallanHarvey85, AlvarezGaumeWitten85, Landsteiner14, GoothEtAl17}. The structure of the CS term reflects the Callan-Harvey mechanism of anomaly cancellation \cite{CallanHarvey85}, provided here by the fermion zero modes living on dislocations/sample boundaries \cite{RanEtAl09,TeoKane2010,Slager2014}. In this way, the CS response related to the 3+1d QHE is valid in the presence of deformations and satisfies the consistency conditions of gauge invariance and anomaly inflow. The same mechanism works for gapped crystalline systems in all odd spatial dimensions. 

The rest of this paper is organized in the following way. Before introducing the elasticity tetrads and the ensuing effective response in 3+1d in Sections \ref{sec:elasticity} and \ref{sec:3DQHE}, we first review the simpler 2+1d QH case. Section \ref{sec:anomaly_inflow} describes the anomaly inflow mechanism of the action in 3+1d. In Sect. \ref{sec:anyD}, we briefly describe the extension to arbitrary even space-time dimensions. The coupling of the elasticity tetrads to spacetime geometry in QHE is discussed in Sec. \ref{sec:gravity} along with the speculative possibility to have quantized and dimensionless gravitational couplings if the gravitational spacetime metric is identified with the metric of elasticity tetrads. We conclude in Sec. \ref{sec:conclusions}.

\section{2+1d topological action for QHE}\label{eq:2DQHE}

The topological Chern-Simons action for the  IQHE and for the anomalous, intrinsic (i.e. without external magnetic field) AQHE  in the $D=2+1$-dimensional crystalline insulator is given by \cite{Ishikawa1986,Volovik1988,Haldane1988,KaplanEtAl93}
\begin{align}
S_{\rm 2+1d}[A_{\mu}]=\frac{1}{4\pi}  N \int d^2 x dt~ \epsilon^{\nu\alpha\beta} A_\nu \partial_\alpha A_\beta\,.
\label{3Daction}
\end{align}
where $|e|=\hbar=1$ and the electromagnetic U(1) gauge field $A_{\mu}$ has dimensions of momentum.
The integer prefactor $N$ in the response is expressed in terms of momentum space topological invariant \cite{Thouless1982, Ishikawa1986,Volovik1988,KaplanEtAl93} 
\begin{align}
N&=\frac{1}{8 \pi^2}\epsilon_{ij} 
\nonumber
\int_{-\infty}^{\infty} d\omega\int_{\rm BZ} dS \\
&~\times {\rm Tr} [(G\doo_{\omega} G^{-1}) (G\doo_{k_i} G^{-1}) ( G\doo_{k_j} G^{-1})]\,,
\label{Invariant}
\end{align}
where the spatial momentum integral is over the 2d torus of the two-dimensional Brillouin zone (BZ). The integer $N$ is a topological invariant of the system and in particular remains locally well-defined under smooth deformations of the lattice. Under sufficiently strong deformations or disorder one can have regions of different $N(x)$ with the associated chiral edge modes. In that case, the global invariant, if any, is defined by the topological charge of the dominating  cluster which percolates through the system \cite{Volovik2018}.

\subsection{2+1d Bulk Chern-Simons and consistent boundary anomaly}\label{sec:2D_anomaly}

The topologically protected physics of the quantum Hall effect arises due to the Callan-Harvey 
\cite{CallanHarvey85} anomaly inflow of the Chern-Simons action from the bulk to the boundary 
\cite{Wen90,Jackiw84,Hughes2013}, which we now review. In fact, the boundary current $J_{\rm bdry}^{\mu}$ realizes the (consistent) 1+1d chiral anomaly
\begin{align}
\doo_{\mu}J_{\rm bdry}^{\mu} = \frac{N}{8\pi} \epsilon^{\mu\nu}F^{\rm bdry}_{\mu\nu} \,,
\end{align}
where $J_{\rm bdry}^{0,\parallel}, F^{\rm bdry}_{0\parallel} = \doo_t A_{\parallel} -  \doo_{\parallel} A_t$ are along the boundary. In this way the protected edge modes arise from the cancellation of bulk and boundary gauge anomalies \cite{Wen90, Jackiw84, CallanHarvey85, KaplanEtAl93}. In more detail, the 2+1d Chern-Simons term Eq. \eqref{3Daction} can be written as
\begin{align}
S_{\rm CS}[A] = \frac{1}{4\pi} \int d^3x \epsilon^{\mu\nu\lambda} \sigma(x) A_{\mu}\doo_{\nu}A_{\lambda}
\end{align}
where $\sigma(x) = N\Theta(x^1)$ has a step function domain wall at $x^1=0$. This is equivalent to a spacetime manifold with a boundary. The current is
\begin{align}
j_{\rm H}^{\mu} = \frac{\delta S_{\rm CS}[A]}{\delta A_{\mu}} = \frac{\sigma}{2\pi} \epsilon^{\mu\nu\lambda} \doo_{\nu}A_{\lambda} - \frac{1}{4\pi} \epsilon^{\mu\nu\lambda}(\doo_{\nu}\sigma) A_{\lambda}
\end{align}
with bulk and boundary contributions. We see that the Hall conductivity is $\sigma_H = Ne^2/h$. The current has the divergence 
\begin{align}
\doo_{\mu}j_{\rm H}^{\mu} = \frac{1}{4\pi} \epsilon^{\mu\nu\lambda}\doo_{\mu}\sigma \doo_{\nu}A_{\lambda} =  \frac{N}{8\pi}\delta(x^1) \epsilon^{1\nu\lambda} F_{\nu \lambda} = \frac{N}{4\pi}E^{\parallel} .
\end{align}
where we have specialized to the case when $F_{02} = -E^{\parallel}$ is an electric field on the boundary. This arises because of the \emph{consistent} gauge anomaly under gauge transformations $\delta_{\lambda}A_{\mu} = \doo_{\mu} \lambda$,
\begin{align}
\delta_{\lambda}S_{\rm CS}[A] = -\frac{1}{4\pi} \int d^3 x~ \lambda \epsilon^{\mu\nu\lambda} \doo_{\mu}\sigma \doo_{\nu}A_{\lambda}  \nonumber\\
= \int d^3x~ \lambda\doo_{\mu}j_H^{\mu} = -\delta_{\lambda}S_{\rm bndry}[A] .
\end{align}
The anomalous divergence is compensated by protected edge modes on the boundary described by $S_{\rm bndry}$.

\section{Elasticity tetrad fields}\label{sec:elasticity}
The pure 3+1d CS term cannot be defined in even spacetime dimensions, therefore the QH response requires additional fields and constitutes a mixed response which is only weakly topologically protected. These fields are due to the (weak) crystalline symmetries of the system. 

Specifically, these are the elastic deformations of the crystal lattice described in terms of elasticity tetrads $E^{~a}_\mu(x)$, which represent the hydrodynamic variables of elasticity theory \cite{DzyalVol1980,VolovikDotsenko1979}. In the absence of dislocations, the tetrads $E^a = E^{\ a}_{\mu} dx^{\mu}$ are exact differentials. They can be expressed in general form in terms of a system of three deformed crystallographic coordinate planes, surfaces of constant phase $X^a(x)=2\pi n^a$, $n^a \in \mathbb{Z}$ with $a=1,2,3$ in three dimensions. The intersections of the three constant surfaces
\begin{equation}
X^1({\bf r},t)=2\pi n^1 \,\,, \,\,  X^2({\bf r},t)=2\pi n^2 \,\,, \,\, X^3({\bf r},t)=2\pi n^3 \,,
\label{points}
\end{equation}
are points of the (possibly deformed) crystal lattice 
\begin{align}
L = \{ \mathbf{r} = {\bf R}(n_1,n_2,n_3) \vert \mathbf{r}\in \mathbb{R}^3, n^a \in \mathbb{Z}^3\}. \label{eq:lattice}
\end{align}
The elasticity tetrads are gradients of the three U(1) phase fields $X^a$, $a=1,2,3$,
\begin{equation}
E^{~a}_\mu(x)= \partial_\mu X^a(x)\,
\label{reciprocal}
\end{equation}
and have units of crystal momentum. By the inverse function theorem, we can define the inverse vectors
\begin{align}
E^{~a}_{\mu}(x) E_{~a}^{\nu}(x) = \delta^{\mu}_{\nu}.
\end{align}
In the simplest undeformed case, $X^a(\mathbf{r},t) = \mathbf{K}^a \cdot \mathbf{r}$, where $E_i^{(0)a} \equiv \mathbf{K}^a$ are the (primitive) reciprocal lattice vectors $\mathbf{K}^a$. In the general case, they depend on space and time and are quantized in terms of the lattice $L$ in Eq. \eqref{eq:lattice}.

In the absence of dislocations, when $X^a(x)$ are globally well-defined, the tetrads $E^{~a}_\mu(x)$ are pure gauge and satisfy the integrability condition (in differential form notation): 
\begin{equation}
T^a = d E^a = \frac{1}{2}(\partial_\mu E^{~a}_\nu(x)-\partial_\nu E^{~a}_\mu(x)) d x^{\mu} \wedge d x^{\nu}=0.
\label{integrability}
\end{equation}
In the presence of dislocations, $T^a \neq 0$, and $X^a(x)$ are multivalued. 

We can also consider a metric associated with these tetrads,
\begin{align}
g_{\mu\nu} = E^{\ a}_{\mu} E^{\ b}_{\nu} \eta_{ab}, \label{eq:lattice_metric}
\end{align}
where $\eta$ is the metric associated to the background lattice, say the spatial Euclidean or Minkowski metric. The important difference is that $dn^2 = g_{\mu\nu} dx^{\mu} dx^{\nu}$ in terms of the elasticity tetrads is dimensionless and therefore counts the spacetime distances in terms of the lattice points of $L$ \cite{AndreevKagan84}. The tetrads fields $E_{\mu}^{\ a}$, but not the metric, enter the CS action for a 3+1d quantum Hall effect of the lattice QH systems and topological insulators as we now discuss. 

\section{3+1d topological action for QHE}\label{sec:3DQHE}

Using the elasticity tetrads, the generalization of the 2+1d QH response to 3+1d is in principle straightforward. 
This extends the electromagnetic 2+1d CS response to $3+1$d with the following topological terms, featuring U(1) fields in combination with the elasticity tetrads $E_{\mu}^{\ a}(x)$:
\begin{align}
S_{\rm 3+1d}[A_{\mu}]=\frac{1}{8\pi^2} \sum_{a=1}^3N_a  \int d^4 x~ E^{~a}_{\mu} \epsilon^{\mu\nu\alpha\beta} A_\nu \partial_\alpha A_\beta\,,
\label{4Daction2}
\end{align}
where $a=1,2,3$ labels the spatial lattice directions in three space dimensions. The derivation of this formula using semi-classical expansion is in the next section.
The tetrads $E_{\mu}^{\ a}(x)$ in Eq. \eqref{reciprocal} have the dimension of the momentum, and thus the integrals in Eq. \eqref{4Daction2} are dimensionless ($\hbar=1$). It follows, as in the 2+1d case, that the prefactors are dimensionless and are also expressed in terms 
of integer topological momentum-space invariants. 
The integer coefficients $N_a$ are antisymmetric integrals of the Green's functions:
\begin{align}
N_a&=\frac{1}{8\pi^2}\epsilon_{ijk} 
\nonumber
\int_{-\infty}^{\infty} d\omega\int_{\rm BZ} dS_a^i\\
&\times~{\rm Tr} [(G\doo_{\omega} G^{-1}) (G\doo_{k_i} G^{-1}) ( G\doo_{k_j} G^{-1})]\,,
\label{Invariants}
\end{align}
where the momentum integral is now over the restricted 2D BZ torus determined by the area $dS^i_a$ normal to $E_{i}^{\ a}$. 

A similar expression for the 3+1d QH was proposed in Ref. \onlinecite{Zubkov16} without the elasticity tetrads and spacetime dependence. In the deformed crystalline systems in 3+1d, the tetrads $E^{~a}_{\mu}(x)$ entering the 3+1d Chern-Simons action Eq. \eqref{4Daction2} depend slowly on space and time. For arbitrary background fields, the spacetime dependence violates the gauge invariance of the action. However, in the absence of dislocations the latter does not happen for the elasticity tetrads: under deformations Eq. (\ref{4Daction2}) remains gauge invariant due to the condition $dE^a=0$ in Eq. \eqref{integrability}.  The variation $\delta_{\phi} S$  under $\delta A_{\mu} = \partial_{\mu} \phi$ is identically zero modulo the bulk/boundary QH currents of the sample. The anomaly cancellation in the presence of dislocations is discussed in detail below. 

Here is the main difference between the topological insulator and a gapless system in 3+1d, e.g. a Weyl semimetal. Integrating out the fermions leads to the chiral anomaly in the bulk and the ensuing gauge anomaly has to be consistently canceled. In that case, instead of the elasticity tetrads the four-momentum separation of the Weyl points appears in the counter term
 \cite{KlinkhamerVolovik05, Haldane, Landsteiner14,CortijoEtAl15, RamamurthyHughes15, GrushinEtAl16, PikulinEtAl16}. In the deformed system, this parameter depends on the coordinates, but the integrability condition is now absent, and the term cannot arise without the bulk chiral gauge anomaly. 
 
\subsection{Semiclassical expansion}\label{sec:semiclassical}

The response in Eq. (8) can be obtained via the semi-classical Green function formalism \cite{Ishikawa1986,Volovik1988,VayrynenVolovik11}. The assumptions are gauge invariance and semiclassical expansion to lowest order. The effective action for a slowly varying background field $A_{\mu}(x,t)$ is
\begin{align}
\log \frac{Z[A]}{Z_0} = \im S_{\rm eff} [A] = \im \tr \ln \frac{G[A]}{G_0} .
\end{align}

Now we want to obtain the effective action to first order in gradients of $A_{\mu}(x)$. We use the fact that the coordinate dependence of $A_{\mu}(x)$ is semiclassical, i.e. the field varies slowly on the scale of the lattice momenta. In the semiclassical phase-space analysis \cite{Ishikawa1986, Volovik1988, VayrynenVolovik11, Volovik2003}, with $G\equiv G[A]\equiv G(\mathbf{p},\omega;A_{\mu}(\mathbf{x},t))$, we obtain to the lowest order in the gradients of $A_{\mu}(x)$
\begin{align}
S_{\rm eff}[A] = 
&-\frac{\im}{4} \int \frac{d^3\mathbf{p} d\omega}{(2\pi)^4} \int d^3\mathbf{x} dt \\
&\tr [G\doo_{x^{\mu}}G^{-1} G \doo_{k_{\mu}}G^{-1} - G\doo_{k_{\mu}}G^{-1}G\doo_{x^{\mu}}G^{-1} \nonumber\\
&\times (G\doo_{k_{\nu}}G^{-1})_{\vert_{A=0}}] A_{\nu} \nonumber
\end{align}
where $x^{\mu} = (t, \mathbf{x})$ and $k_{\mu} = (\omega, -\mathbf{p})$ and
\begin{align}
G\doo_{x^{\mu}}G^{-1} = G\doo_{k_{\nu}}G^{-1}_{\vert_{A=0}}\doo_{x^{\mu}}A_{\nu}.
\end{align}
The trace in the matrix product of Green functions is antisymmetric in the indices of $k_{\mu}$-derivatives. 
With the convention $\epsilon^{txyz} = -1$, we arrive to
\begin{align}
S_{\rm eff}[A]&= \frac{\im}{12}  \int d^3 x dt~ \epsilon^{\alpha\beta\gamma \rho} A_{\alpha}\doo_{\beta}A_{\gamma}~ \times \\  
\int \frac{d^3\mathbf{p}d\omega}{(2\pi)^4} &
\epsilon_{\mu\nu\lambda \rho}\tr[(G\doo_{k_{\mu}}G^{-1})(G\doo_{k_{\nu}}G^{-1})(G\doo_{k_{\lambda}}G^{-1})]_{A=0} \nonumber.
\end{align}
The momentum space prefactor can be separated to be of the form
\begin{align}
\im \int d p^a \delta_a^{\rho} \epsilon_{\mu\nu\lambda\rho} \int & \frac{d^2 \mathbf{p} d\omega}{24\pi^2} \times \nonumber\\
\tr[(G\doo_{k_{\mu}}G^{-1}) & (G\doo_{k_{\nu}}G^{-1})(G\doo_{k_{\lambda}}G^{-1})]_{A=0} \\
= \int d p^a N_a(p^a) &= K^a N_a \nonumber,
\end{align}
where $N_a = N_a(p_a)$ is the 2+1d integer-valued invariant,
\begin{align}
N_a(p^a) &=  \epsilon_{\mu\nu\lambda a} \int \frac{d^2 \mathbf{p} d\omega}{24\pi^2} \times  \nonumber\\ 
&\tr[(G\doo_{k_{\mu}}G^{-1}) (G\doo_{k_{\nu}}G^{-1})(G\doo_{k_{\lambda}}G^{-1})]_{A=0} 
\end{align}
evaluated in imaginary time over the cross-sectional reciprocal space, perpendicular to the reciprocal lattice direction $K^a$. The elementary cell can be taken to be triclinic. Topologically the integration is over a (pinched) 3-torus and $N_a(p^a)$ is an element of $\pi_3(GL(n,\mathbb{C})) = \mathbb{Z}$ \cite{Ishikawa1986}.

We conclude that
\begin{align}
S_{\rm eff}[A] = \frac{1}{8\pi^2} N_a \int d^3 \mathbf{x} dt~ \epsilon^{\mu\nu\lambda\rho} E^{(0)a}_{\ \mu}  A_{\nu}\doo_\lambda A_{\rho} .
\end{align}
where $E^{(0)a}_{\ \mu} = (K^a)_i \delta^i_{\mu}=\int dp^a$, $i=x,y,z$ are the reciprocal lattice vectors normal to the different lattice planes with invariants $N_a$. This is the statement that the 3+1d QHE is described by the weak vector invariant $N_a E^{(0)a}_{\mu}$ protected by the crystalline symmetry.

It is well-known that the topological winding number $N_a$ is stable against small variations $\delta G(k_x,k_y,k_z)$ of the Green's function that do not close the gap 
\cite{Ishikawa1986,Volovik1988,Volovik2003}. 
We can consider small lattice deformations $x^{\mu} \to x^{\mu}+\xi^\mu(x)$, $\xi\ll 1$. Under these, the reciprocal vectors change as
\begin{align}
E^{(0)a}_{\ \mu} &= \int dp^a \to   \nonumber \\ 
E^{\ a}_{\mu}(x) 
&\equiv \int dp^a(x) \approx \frac{2\pi}{d}(\delta^a_{\mu}-\doo_{\mu}\xi^a) 
\end{align}
where $E^{\ a}_{\mu}(x)$ is a semiclassical, slowly varying field at the lattice scale $d$ in reciprocal space which defines the local normal direction of a set of lattice planes. Note that the topological invariants $N_a$ are assumed to be constant and independent of deformations throughout. The final result is Eq. (8).

\subsection{Hall current in terms of elasticity tetrads}

The elasticity tetrads, i.e., elementary deformed reciprocal and direct lattice vectors, $E^{\ a}_{\mu}(x) = \doo_{\mu}X^a(x)$ and the inverse $E_a^{\ \mu}(x)$, appear in the EM response, descending from the lattice field phase fields $N_a X^a(x)$ in the presence of deformarions. As discussed, the phase fields satisfy the quantization condition $X^a(x) = 2\pi n^a$, $n^a\in \mathbb{Z}^3$, on the lattice points $x\in L$. The Hall conductance is \cite{Halperin1992}
\begin{align}
\sigma_{ij} = \epsilon_{ijk} \frac{\sum_a N_a E^{a}_{\ k}(x)}{4\pi^2},
\end{align}
i.e., the conductance is quantized in planes perpendicular to the layer normal $G_i = \sum_a N_a E^a_{\ i}(x)$. The reciprocal vector $\mathbf{G} = G_i$ is the weak Hall index of a weak 3+1d Chern insulator.

In the non-deformed crystal, where  $E^{(0)a}_{k}$ are primitive reciprocal lattice vectors, this equation transforms to the well-known equation with 3+1d quantized Hall 
conductivity \cite{Halperin1987,Halperin1992,Haldane,KlinkhamerVolovik05}:
\begin{equation}
\sigma_{ij}=\frac{e^2}{2\pi h}\epsilon_{ijk}G_k\,,
\label{ConductivityHomog}
\end{equation}
where $G_k$ is a reciprocal lattice vector, which is expressed in terms of the topological invariants $N_a$ and the primitive reciprocal lattice vectors $E^{(0)a} = \mathbf{K}^a$:
\begin{equation}
G_k=\sum_{a=1}^3 N_a E^{(0)a}_k\,.
\label{ReciprocalVector}
\end{equation}

In the deformed crystal, the conductivity tensor is space-time dependent, and thus is not universally quantized. However, the response of the conductivity to deformation is quantized:
\begin{equation}
\frac{d\sigma_{ij}}{dE^{\ a}_k} =\frac{e^2}{2\pi h}\epsilon_{ijk}  N_a\,.
\label{ConductivityVariation}
\end{equation}
The Hall current is
\begin{align}
J ^{\mu} 
&= \frac{-1}{4\pi^2} \sum_{a=1}^3N_a \epsilon^{\mu\nu\alpha\beta}  E^{~a}_{\nu}   \partial_\alpha A_\beta \nonumber \\
&+ \frac{1}{8\pi^2} \sum_{a=1}^3N_a \epsilon^{\mu\nu\alpha\beta}  (\doo_{\alpha}E^{~a}_{\beta}) A_\nu 
\label{Current}
\end{align}
and has a bulk and a topological defect/boundary component \cite{CallanHarvey85, Naculich88}. In the absence of dislocations, $d E^a = 0$ in Eq. \eqref{integrability}, and the current represents a dissipationless, fully reversible current,
which is conserved due to U(1) gauge invariance 
\begin{equation}
\partial_\mu J^{\mu} = \doo_{\mu}J^{\mu}_{\rm bulk}=0.
\label{CurrentConserv}
\end{equation}

\subsection{Chiral magnetic effect}

The action Eq. (\ref{4Daction2}) and current Eq. \eqref{Current} also describe the chiral magnetic effect (CME) \cite{FukushimaEtAl08, LiEtAl16}. In the presence of periodic directions varying in time, time-dependence $X^a(\mathbf{r},t)$ appears. In the CME an electric current along an applied magnetic field is induced:
\begin{align}
{\bf J}=\frac{1}{4\pi^2}\sum_{a=1}^3 N_aE^{\ a}_t\mathbf{B} \,.
\label{conductivity}
\end{align}
This current contains $E^{~a}_t=\partial X^a/\partial t$ and thus it vanishes in equilibrium in agreement with Bloch theorem, according to which the total current is absent in the ground, or in general any equilibrium, state of the system (see, e.g., Ref. \onlinecite{Yamamoto2015}). Here, we restrict to spatial lattices under deformations. The CME for a time-periodic insulator with timelike $E^{ 0}_{\mu} = \omega_F \delta^0_{\mu}$ (with Floquet drive $\omega_F$) and the temporal invariant $N_0 \neq 0$ was pointed out in Ref. \onlinecite{NissinenVolovik2018b}. This can be extended to spatial deformations as well.

\section{Anomaly cancellation and dislocation zero modes}\label{sec:anomaly_inflow}

\subsection{Callan-Harvey effect on dislocations and mixed anomaly}

Now, we describe the anomaly inflow. The constraint \eqref{integrability} is violated in the presence of topological defects  --- dislocations. 
The density of dislocations equals the torsion for the elasticity tetrads (with vanishing spin connection):
\begin{align}
T^a = dE^a, \quad T^a_{kl} =(\partial_k E^{~a}_l-\partial_l E^{~a}_k),\quad k,l = x,y,z,
\label{Torsion}
\end{align}
similar to the role of spacetime torsion in gravitational theories \cite{DzyalVol1980, Kleinert}. The nonzero dislocation density or torsion violates the conservation of the Hall current:
\begin{equation}
\partial_\mu J^{\mu} =\frac{-1}{8\pi^2} \frac{1}{4} \epsilon^{\mu\nu\alpha\beta} F_{\alpha\beta}
\sum_{a=1}^3N_a  T^a_{\mu\nu} 
\,.
\label{CurrentNonconserv}
\end{equation}
This mixed anomaly represents the Callan-Harvey mechanism of anomaly
cancellation \cite{CallanHarvey85}, which is provided here by the fermion zero modes living on dislocations \cite{CallanHarvey85, Naculich88, RanEtAl09, TeoKane2010,Slager2014}. The action remains well-defined and gauge invariant in the presence of dislocations, i.e., $2\pi$ ambiguities in the phase field $X^a$, due to zero modes with 1+1D covariant anomaly along the dislocation string,
\begin{align}
\partial_\mu J^{\mu} = \doo_{\mu}J^{\mu}_{\rm bulk}+\doo_{\mu}J^{\mu}_{\rm dislocation}=0,
\label{CurrentConserv}
\end{align}
as shown in Refs. \cite{CallanHarvey85, Naculich88, WilczekGoldstone81} for Dirac fermions in the presence of complex vortex-like axionic mass. The Dirac model can be taken as a topological model for the 3+1d QH system \cite{Volovik2003}, which we discuss next.

\subsection{Dirac fermion model} \label{app:dislocation}
\begin{figure}
\includegraphics[width=.49\textwidth]{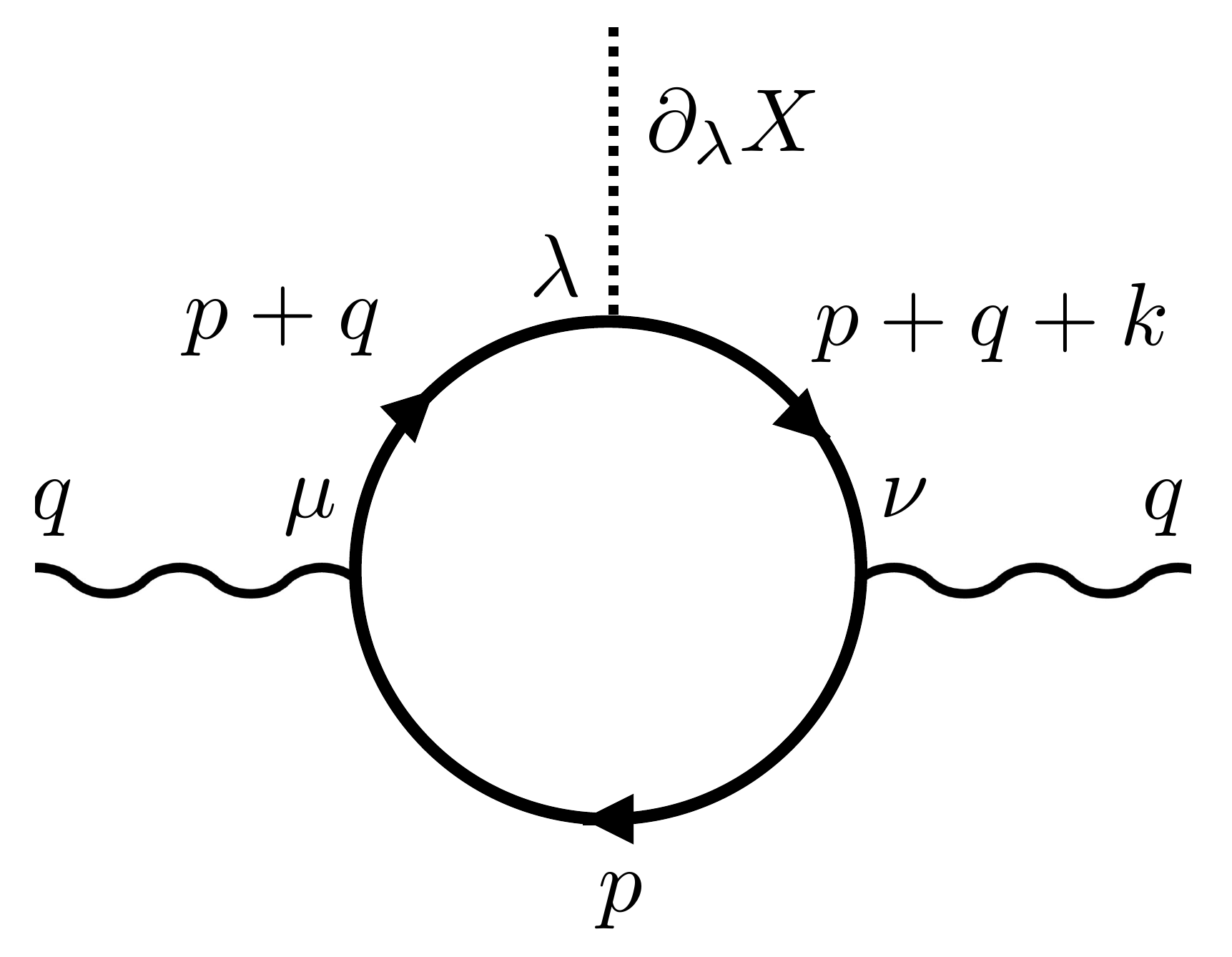}
\caption{The four-dimensional gauge field vacuum polarization $\Pi^{\mu\nu}(q,k)$ in Eq. \eqref{eq:polarization} with the axionic vertex $\doo_{\lambda} X$.}
\label{fig:axion_polarization}
\end{figure}

To verify the existence of dislocation zero modes on dislocations and overall gauge invariance, let us consider a simplified model with the same symmetries as the time-reversal and parity-breaking topological insulator with crystalline order, namely, a gapped and time-reversal and parity breaking Dirac model with a complex mass induced by a scalar field $X$, 
\begin{align}
&S[\psi,\overline{\psi},A_{\mu},X] \nonumber\\
&=  \int d^3 \mathbf{x} dt~ \overline\psi (\gamma^{\mu}\im\doo_{\mu})\psi - m(x) \overline{\psi}e^{\im \gamma^5X}\psi - \overline{\psi}\gamma^{\mu}\psi A_{\mu} \nonumber \\
&= \int d^3 \mathbf{x} dt~ \overline\psi (\gamma^{\mu}\im\doo_{\mu}-m)\psi \nonumber \\ 
&\quad - m \overline{\psi}[e^{\im \gamma^5X}-1]\psi - \overline{\psi}\gamma^{\mu}\psi A_{\mu}+ \cdots  .\label{eq:axion_string}
\end{align}  
The complex scalar $m(x)e^{\im \gamma^5 X(x)}$ is equivalent to a slowly varying mass profile. A vortex line-singularity (``axionic string" \cite{CallanHarvey85, Naculich88}) is the profile where $m(x) \to m$ far from the singularity and $m(x_c) = 0$, where $x_c$ is the vortex location. The phase field $X(x)$ has a vortex singularity
\begin{align}
\int_C dX = 2\pi n \label{eq:vortex_number}
\end{align}
around all contours around $x_c$ and is is such that $X=\textrm{const.} + n\phi$ far from the line defect. In the present case, $X= N_a X^a$ and $n = n^aN_a$. The expansion in Eq. \eqref{eq:axion_string} applies far from the vortex. Let us compute the induced effective action \emph{far} from the vortex string by starting the from the expansion
\begin{align}
m \overline{\psi}[e^{\im \gamma^5X}-1]\psi (y) = \im m \doo_{\mu} X (y-x)^{\mu} \overline{\psi}\gamma^5 \psi(x) + \dots
\end{align}
The response is found from the axial bosonic polarization vertex $\Pi^{\mu\nu}(x,y)$ \cite{Naculich88} (or the equivalent Goldstone-Wilczek current \cite{WilczekGoldstone81}) of the effective action
\begin{align}
S_{\rm eff}[A] = \int d^4x d^4y~ A_{\mu}(x) \tfrac{1}{2}\Pi^{\mu\nu}(x,y) A_{\nu}(y)
\end{align}
depicted in Fig. \ref{fig:axion_polarization} becomes, using $\tr[\gamma^5\gamma^{\mu}\gamma^{\nu}\gamma^{\lambda}\gamma^{\rho}] = +4\im\epsilon^{\mu\nu\lambda\rho}$ with $\epsilon^{txyz}=-1$,
\begin{widetext}
\begin{align}
\Pi^{\mu\nu}(x,y) =  m \doo_{\lambda}X& \int \frac{d^4 q d^4 p}{(2\pi)^8}d^4k \delta^{(4)}(k)  e^{\im q(y-x)}\frac{\doo}{\doo k_{\lambda}}\tr [\im\gamma^{\mu} G(p)\im \gamma^{\nu}G(p+q)\gamma^5G(p+q+k) \nonumber\\
& \phantom{XXXXXXXXXXXXXZZZZZZZZZZZZZZZ}+ (\mu\leftrightarrow \nu, k\leftrightarrow q)] \label{eq:polarization}
\\=+4\im \epsilon^{\mu\nu\lambda\rho} \doo_{\lambda}X &\int \frac{d^4 q d^4 p}{(2\pi)^8}  e^{\im q(y-x)}\frac{m^2 \im q_{\rho}}{(p^2-m^2)((p+q)^2-m^2)}\left[\frac{1}{p^2-m^2}+\frac{1}{(p+q)^2-m^2}\right] \nonumber.
\end{align}
\end{widetext}
Taking the limit $q\to 0$, we arrive to the \emph{bulk} effective action \cite{CallanHarvey85, Naculich88}
\begin{align}
S_{\rm eff}[A,X] = \sum_{a} N_a \frac{1}{8\pi^2}\int d^4x~  \epsilon^{\mu\nu\lambda\rho} \doo_{\mu}X^a A_{\nu}\doo_{\lambda}A_{\rho} + \cdots \label{eq:bulk_current}
\end{align}
after integrating out the massive fermions \emph{far} from the string. The response follows from Eq. \eqref{eq:polarization} by the replacement $X \to \sum_a N_a X^a$, i.e. instead of a single U(1) axion field, we have a flavored U(1)$^3$ lattice phase field $X^a$ with topological charges $N_a$ which are \emph{not} protected \cite{VayrynenVolovik11} without the existence of the slowly varying lattice phase field $X^a$.

Let us now extend the above effective action Eq. \eqref{eq:bulk_current} to be valid \emph{everywhere}, essentially by considering a vortex string with a delta function core. 

\subsection{Induced anomaly current} \label{app:anomaly}
The bulk effective action Eq. \eqref{eq:bulk_current} was derived far from the string defect. Let us now check the induced current due to a topological configuration in $X(x)=N_a X^a(x)$. For simplicity, we will consider singularities only in the slowly varying continuum lattice field $X^a(x)$ and treat the topological numbers $N_a$ as constants. A dislocation in the topological response is a $2\pi n^a N_a$-ambiguity in the field $X = N_a X^a(x)$, and $n^a$ is the Burgers vector of the dislocation and $n=n^a N_a$. 

The induced current separates to a bulk and boundary contributions,
\begin{align}
\langle J^{\mu} \rangle = \delta S_{\rm eff}/\delta A_{\mu}  = \frac{-1}{4\pi^2} \epsilon^{\mu\nu\lambda\rho}\doo_{\nu} X \doo_{\lambda}A_{\rho} \nonumber\\
+\frac{1}{8\pi^2} \epsilon^{\mu\nu\lambda\rho}(\doo_{\lambda}\doo_{\nu} X) A_{\rho}. \label{eq:induced_bulk_current}
\end{align}
due to an electric field $F_{0z} = -E^z$ parallel along the string. This has contributions from the bulk, where $X(r,\phi,z)=n\phi$, and from the string where, from Eq. \eqref{eq:vortex_number}, we get
\begin{align}
-\epsilon^{tijz}\doo_i \doo_j X = 2\pi n \delta^{(2)}(x_c) ,\quad i,j=x,y.
\end{align}
Therefore,
\begin{align}
\langle J^{\mu}_{\rm bulk}\rangle &=\frac{-1}{4\pi^2} \epsilon^{\mu\nu\lambda\rho}\doo_{\nu} X \doo_{\lambda}A_{\rho} \\ 
&= \frac{1}{4\pi^2}\frac{n}{r} F_{0z}  \nonumber = -\frac{1}{4\pi^2} \frac{n}{r} E^z = \langle J^r_{\rm bulk} \rangle,
\end{align}
whence the current per unit time and length in the radial direction outwards from the string is $-\frac{n}{2\pi}E^z$. In addition, there is the current localized on the string
\begin{align}
\langle J_{\rm string}^{i} \rangle &= \frac{1}{8\pi^2} \epsilon^{\mu\nu\lambda\rho}(\doo_{\lambda}\doo_{\nu} X) A_{\rho} \\
&= \frac{1}{8\pi^2}(2\pi n \delta^{(2)}(x_c))\epsilon^{ij}A_j = \frac{n}{4\pi} \epsilon^{ij}A_j \delta^{(2)}(x_c), \nonumber
\end{align}
$i,j=t,z$, with the divergence $\doo_i \langle J_{\rm string}^i \rangle = \frac{n}{8\pi} \epsilon^{ij}F_{ij} \delta^{(2)}(x_c)  = \frac{n}{4\pi} E^z \delta^{(2)}(x_c)$.
The divergence of the total current Eq. \eqref{eq:induced_bulk_current} on the string is anomalous
\begin{align}
\doo_{\mu}\langle J^{\mu} \rangle = -\frac{1}{8\pi^2}\epsilon^{\mu\nu\lambda\rho} (\doo_{\mu} \doo_{\nu} X) F_{\lambda\rho} \\
= \frac{1}{4\pi^2} (2\pi n \delta^{(2)}(x_c) )F_{0z} = -\frac{n}{2\pi}E^z \delta^{(2)}(x_c). \nonumber
\end{align}
On the other hand the chiral fermions on the string, where they are effectively massless, result in $n$ chiral zero modes with chiral anomaly
\begin{align}
\doo_{i} \langle J_{\rm 1+1D}^{i} \rangle = \doo_i \langle J_{\rm string}^{i}\rangle + \doo_{i} \langle J^{i, \rm cons.}_{\rm 1+1D} \rangle \\
= \pm \frac{1}{4\pi} \epsilon^{ij} F_{ij} = \mp \frac{F_{0z}}{2\pi} = \pm \frac{E^z}{2\pi}. \nonumber
\end{align}
This is the 1+1d \emph{covariant} anomaly and is composed of the boundary consistent anomaly plus the contribution on localized on the string from the bulk \cite{Naculich88, Landsteiner14}. The anomaly due to the bulk induced current and zero modes cancels.

In contrast, from the previous arguments in Sec. \ref{sec:2D_anomaly} we see that the 2+1d QH anomaly is matched by the \emph{consistent} anomaly of $N$ boundary 1+1d chiral fermions. This seems to coincide with the observation in Ref. \onlinecite{Landsteiner14}:  The \emph{covariant} anomaly is a Fermi-surface effect, whereas the \emph{consistent} anomaly arises from the contribution of all the filled levels.

\section{Extension to even $D=2+2n$ space-time dimensions}\label{sec:anyD}

The anomaly equation Eq. \eqref{CurrentNonconserv} can be straightforwardly extended to arbitrary even $D=2+2n$ spacetime dimensions:
\begin{equation}
\partial_\mu J^{\mu}  \propto  
\sum_{a=1}^{2n+1}N_a  T^a \wedge \underbrace{F  \wedge ... \wedge F}_{n \textrm{ times}}
\,.
\label{CurrentNonGener}
\end{equation}
In addition to the torsional field strength $T^a = dE^a$, it contains the $n$-fold antisymmetric tensor product of U(1) gauge field strengths $F = dA$, while the integer valued topological invariants $N_a$ are expressed in terms of $2n+1$-dimensional integrals in frequency-momentum space \cite{KaplanEtAl93}.

Equation \eqref{CurrentNonGener} is valid even in case of $n=0$, i.e., in one spatial dimension, producing the expected results.
Consider a gapped one-dimensional chain of electrons with the action (see also \cite{RamamurthyHughes15})
\begin{align}
S_{\rm 1+1d}[A] = \frac{N_1}{2\pi} \int dx dt ~ \epsilon^{\mu\nu} E^{\ 1}_{\mu} A_{\nu},  \label{eq:action_1+1D}
\end{align}
where the index $N_1$ is defined via the Green's function $G(\omega,k_x)$,
\begin{align}
N_1 &= \frac{1}{2\pi\im} \int d\omega \textrm{ Tr } G(k_x,\omega) \doo_{\omega}G^{-1}(k_x,\omega) \,.
\label{eq:N_1}
\end{align}
The index $N_1$ is the same for any $k_x$ without gap closings. Note that in 3+1d the index $N_a$ in 
Eq.(\ref{Invariants}) is the same for any  cross section $\mathbf{S}_a$ of the three-dimensional Brillouin zone, while in 1+1d the cross section corresponds to one point $k_x$ in the one-dimensional Brillouin zone.

Variation of Eq. \eqref{eq:action_1+1D} gives the electric current
\begin{align}
 J^{\mu} = \frac{N_1}{2\pi}\epsilon^{\mu\nu} E^{\ 1}_{\nu} \,,
\label{eq:1Dcurrent}
\end{align}
the conservation of which 
\begin{align}
\doo_{\mu}J^{\mu} = N_a d E^{a} = N_1 dE^{1} = 0 \,,
\label{1Dconservation}
\end{align}
has a simple interpretation.
The condition $dE^a = 0$ is equivalent to the conservation of the sites of the one-dimensional lattice, 
 whereas the index $N_1$  corresponds to the number of the electrons per site, which is integer for band insulators. As a result, the number of the electrons is trivially conserved under adiabatic deformations.

Since any one-dimensional insulator is described by the topological invariant $N_1$, which can only change when the gap closes, we may call any 1+1d insulator topological, although the topology of the filled states can only be detected by higher invariants --- the Chern numbers. In fact, this is very similar to ordinary metals with Fermi surfaces \cite{Volovik2003}: the gapless Fermi surface represents a topological object in momentum space protected by an invariant similar to $N_1$. Topology provides the stability of the Fermi surface with respect to interactions and explains why metals can be described by Landau Fermi-liquid theory. In this sense, metals can be considered as topological materials, making the zeroth-order invariant $N_1$ one of the most important topological invariants in the hierarchy of the topological invariants for fermionic  systems. In particular, it gives rise to the Luttinger theorem: the number of states in the region between two Fermi points in 1+1d does not depend on interaction and arises through topology and adiabatic evolution. This can be generalized for any closed Fermi surface or insulator in higher dimensions \cite{Oshikawa2000, Volovik2003, HeathBedell19}.

\section{Gravitational QH response and emergent gravity}\label{sec:gravity}

Now, we briefly describe the anomalous coupling of the elasticity tetrads to the effective space-time metric in the quantum Hall system and speculate on the possibility of relating the elastic metric with the gravitational space-time metric in an effective low-energy effective model of quantum gravity.

The dimensional elasticity tetrads in the 3+1d QHE allow us to write the dimensional extension of the 2+1d gravitational framing anomaly as a mixed elastic-gravitational anomaly in 3+1d \cite{CallanHarvey85, Witten89, Volovik90,ReadGreen01, AbanovGromov14}:
\begin{align}
S_{\rm eff, g} =  \sum_{a=1}^{3} \frac{\widetilde{N}_a}{192\pi^2} \int d^3x dt~ \nonumber
 E^a \wedge 
\bigg(  \Gamma^\mu_{\ \nu} \wedge d\Gamma^\nu_{\ \mu} \\
+\frac{2}{3}  \Gamma^\mu_{\ \nu} \wedge  \Gamma^\nu_{\ \rho} \wedge  \Gamma^\rho_{\ \mu} \bigg),
\label{CurrentThermal}
\end{align}
where $\widetilde{N}_a$ are effective central charges, 
$\Gamma^{\mu}_{\ \nu} \equiv \Gamma^\mu_{\lambda \nu} dx^{\lambda}$ are the Christoffel symbol one-forms of the space-time metric, that arises for example through Luttinger's argument \cite{Luttinger64}. In particular, the mixed elastic-gravitational Chern-Simons term \eqref{CurrentThermal} implies the generalization of the 2+1d thermal Hall effect \cite{Luttinger64, ReadGreen01, Stone12, Stone2018, Abanov2015} to 3+1d quantum Hall systems and topological insulators with intrinsic Hall effect, although it is third order in derivatives and therefore beyond linear response.

 Finally, let us speculate on the hypothetical connection of the elasticity tetrads $E^{\ a}_{\mu}$ to real space-time metric and gravity. While the microscopic structure of the relativistic quantum space-time vacuum is not known, phenomenological approaches and effective field theory can be used to describe the effects of the vacuum degrees of freedom. In one of these scenarios, it is assumed  that the space-time vacuum has  the properties of a $3+1$d super-plastic crystalline medium constructed from $E^{\ a}_{\mu}$ \cite{KlinkhamerVolovik2019}.  As in the condensed matter elasticity theory, dislocations and disclinations in this space-time crystal describe torsion and curvature of general relativity \cite{Bilby1956,Kroener1960,DzyalVol1980,KleinertZaanen2004}.
More specifically, in this super-plastic model of gravity, the size of the elementary cell in the vacuum space-time crystal is not fixed but in principle can vary arbitrarily with no elementary Planck scale space-time lattice. As a result, the induced action for the gravitational field \cite{KlinkhamerVolovik2019}
\begin{equation}
S[E^a_{\ \mu}]=
\int_{M^{3,1}}
\,d^4x\, \vert E \vert\,
\left(KR+\Lambda
\right) \,,
\label{eq:action-q}
\end{equation}
contains only dimensionless quantities. Here the metric $g_{\mu\nu} = \eta_{ab}E^a_{\ \mu}E^b_{\ \nu}$ is from Eq. \eqref{eq:lattice_metric} and the Ricci curvature scalar $R[E^a_{\ \mu}]$ depends on the elasticity tetrads in the standard way, thus making the gravitational constant $K$ (inverse Newton constant) and the cosmological constant $\Lambda$ dimensionless, $[K]=[R]=[\Lambda]=1$. The same holds for
higher derivative gravity and for the other (non-gravitational) physical quantities, such as particle mass $[M]=1$. Therefore, if gravity is related to fundamental elasticity tetrads, all the measurable physical quantities are dimensionless with respect to the space-time lattice.

\section{Conclusion}\label{sec:conclusions}

In this paper, we have described the mixed topological response of 3+1d QH systems using elasticity tetrads. We have shown how gauge invariance and anomaly inflow arise in the presence of deformations of the weak lattice symmetries protecting the state. The response is a mixed CS response featuring the elasticity tetrads fields of the lattice. This mixed and geometric nature of the 3+1d CS response is to be contrasted with the more familiar topological BF theory in four dimensions (see, e.g., Ref. \onlinecite{ChoMoore11}), which represents a mixed response with two one-form gauge fields and their the field strengths appearing in the action.

More generally, the dimensionful elasticity tetrads are the proper hydrodynamic variables related to the weakly protected IQHE/AQHE/CME on general even-dimensional crystalline backgrounds. They describe the topological QH response with elastic and geometric deformations in systems in odd spatial dimensions. In general, Eq. \eqref{CurrentNonconserv} and its extension to higher dimensions, Eq. \eqref{CurrentNonGener}, describe the  mixed anomaly in terms of the gauge fields and elastic torsion.
These equations do not contain any parameters, except for topological quantum numbers. This is because the elastic tetrads (torsion) have the canonical dimensions of $[l]^{-1}$ (respectively $[l]^{-2}$) instead of the conventional $l^0$ (respectively $[l]^{-1}$) for the gravitational space-time tetrads (torsion). This allows one to write many mixed ``quasi-topological Chern-Simons terms", analogous to mixed axial-gravitational/elastic anomalies, in 3+1d quantum Hall systems with weak crystalline symmetries. In principle, we can envisage similar geometric extension of BF theory to 4+1d using elasticity tetrads.

{\bf Acknowledgement}. This work has been supported by the European Research Council (ERC) under the European Union's Horizon 2020 research and innovation programme (Grant Agreement No. 694248).

\end{document}